\documentclass[12pt]{JHEP3}
\usepackage{amsmath,amssymb,epsfig,graphicx}

\newcommand{\be}{\begin{equation}}
\newcommand{\ee}{\end{equation}}
\newcommand{\beqa}{\begin{eqnarray}}
\newcommand{\eeqa}{\end{eqnarray}}

\def\eeq{\end{equation}}

\relax

\def\beq{\begin{equation}}
\def\eeq{\end{equation}}
\def\bea{\begin{eqnarray}}
\def\eea{\end{eqnarray}}

\title{Constraint fitting of experimental data with a jet quenching model embedded in a hydrodynamical bulk medium}

\author{N\'estor~Armesto${}^{\,a}$, Matteo Cacciari${}^{\,b}$, Tetsufumi~Hirano${}^{\,c}$, James L. Nagle${}^{\,d}$
and Carlos~A.~Salgado${}^{\,a,\, b}$\\
\vspace{0.1in}

${}^{\,a}$Departamento de F\'\i sica de Part\'\i culas and IGFAE,
Universidade de Santiago de Compostela, E-15706 Santiago de Compostela, Galicia--Spain
\vspace{0.1in}

${}^{\,b}$ LPTHE, UPMC -- Paris 6, Paris-Diderot -- Paris 7, CNRS UMR 7589, Paris, France
\vspace{0.1in}

${}^{\,c}$ Department of Physics, The University of Tokyo, Tokyo 113-0033, Japan
\vspace{0.1in}

${}^{\,d}$ University of Colorado,
Boulder, CO 80309-0390, USA
\vspace{0.1in}

E-mail addresses: {\tt nestor.armesto@usc.es, cacciari@lpthe.jussieu.fr, hirano@phys.s.u-tokyo.ac.jp, jamie.nagle@colorado.edu, carlos.salgado@usc.es}
}

\abstract{
We present a global fit to single- and double-inclusive suppression data of high-$p_T$ particles in central Au+Au collisions at top RHIC energy.  We also include in this analysis data on heavy quarks via their D and B meson semi-leptonic decays (i.e. non-photonic electrons).  The analysis is based on the parton quenching weights for medium-induced gluon radiation computed in the BDMPS approximation then embedded in a hydrodynamical description of the bulk medium. Our results indicate that values of the transport coefficient $\hat q$ more than four times larger than perturbative estimates are preferred by experimental data. This confirms previous calculations based on simpler implementations of the medium geometry or only the single-inclusive suppression.  We also comment on the statistical compatibility of the heavy quark data within a radiative only energy loss scenario, and on the sensitivity of the results to nuclear modification of the parton distribution functions (PDFs) and to assumptions on the energy loss during times in the collision prior to the hydrodynamical behavior.}

\preprint{\today}

\begin{document}

\section{Introduction}

The experimental program at the Relativistic Heavy Ion Collider (RHIC) at Brookhaven National Laboratory (BNL) has established the {\it jet quenching} phenomenon as a fundamental tool in the study of hot matter in heavy-ion collisions~\cite{RHIC} -- see \cite{reports} for reports on the subject. A variety of observables in the large transverse momentum region of the produced particle spectrum is contained under this generic name: the suppression of the single-inclusive yields of both light hadrons and non-photonic electrons; the suppression of the double-inclusive azimuthal correlations; direct photon-hadron correlations; the effects on identified jets, jet shapes; etc.  A subset of these observables are also available,
though with different expected magnitude, at the Super Proton Synchrotron (SPS) at CERN~\cite{SPS}, and all of these observables will play a major role in the upcoming heavy ion program at the Large Hadron Collider (LHC) at CERN.

Another essential discovery at RHIC is the good description of the data, at small and moderate transverse momentum, achieved by a hydrodynamical modeling of the matter produced in heavy-ion collisions~\cite{RHIC}.  In particular, azimuthal asymmetries compatible with a hydrodynamical flow starting at very early times and with negligible viscosity have been measured~\cite{hydroreviews}.

Although there is no general consensus, several jet quenching studies \cite{Dainese:2004te,Eskola:2004cr,Renk:2006pk,Renk:2006sx,Bass:2008rv} require an interaction of partons with the medium which exceeds that estimated from the asymptoticaly large temperature behavior of perturbative Quantum Chromodynamics (pQCD) \cite{Baier:2002tc,Baier:2006fr,CasalderreySolana:2007sw,Baier:2008js}.  
Corrections to these estimates (arising from higher orders in perturbation theory) are not known for quantities as e.g. $\hat q$ but indications exist that they could be sizable \cite{CaronHuot:2008ni}. It is also worth recalling here that the relevant scale in the perturbative expansion, both in the medium and in the vacuum, is the transverse momentum of the radiated gluon. For an opaque medium case, this quantity acquires additive contributions from multiple soft scatterings which can be described in terms of light-like Wilson lines \cite{reports,Liu:2006he}. The averages over medium configurations of these non-perturbative quantities are usually encoded in the single parameter $\hat q$. Subleading terms in the high-energy approximation involved in calculations of radiative energy loss \cite{reports} are not yet known.
Thus, the conclusion of the large strength remains, but caution is warranted on the precise interpretation of the actual magnitude obtained. 

These findings about the large partonic interaction strength in jet quenching and the hydrodynamical behavior of the bulk matter have led to the claim that the produced medium is strongly coupled~\cite{scqgp}. Such a conclusion is of critical importance for our understanding of QCD. Therefore it must be substantiated by detailed studies which should consider: i) As many experimental observables as available; ii) A detailed modeling of the medium compatible with experimental data on soft particle production; iii) A statistical analysis of the uncertainties in the constraints on the medium coming from both the experimental data and the theoretical implementation of energy loss and medium modeling.

The goal of the present paper is to constrain the properties of the medium which can be studied through the description of its effects on high-$p_T$ particles.
The medium is described by a hydrodynamical model adjusted to fit experimental data on the soft part of the spectrum. The modeling of the energy loss, considered as purely due to medium-induced gluon radiation, is that proposed some years ago \cite{Baier:1996sk,Zakharov:1998sv,Wiedemann:2000za,Salgado:2003gb} and used in several phenomenological studies of RHIC data \cite{Dainese:2004te,Eskola:2004cr,Bass:2008rv,phenix}. The coupling of a hydrodynamical medium description to an energy loss calculation has already several antecedents \cite{Hirano:2003hq,Hirano:2003pw,Hirano:2003yp,Hirano:2004en,Renk:2006pk,Renk:2006sx,Bass:2008rv}.  Fits to experimental data using $\chi^2$-minimization of statistical uncertainties for single- and double-inclusive data were presented in \cite{Zhang:2007ja}, and including statistical and systematic
uncertainties in~\cite{Nagle:2008fw,phenix}.

In this work, we extend previous studies by both including a hydrodynamical modeling of the medium and considering the case of inclusive pion, dihadron and heavy quark suppression within the same analysis. The energy loss distributions will be taken from \cite{Salgado:2003gb} for light quarks and gluons and from \cite{Armesto:2005iq} for heavy quarks.  The observables related to direct photons-hadron correlations and identified jets have not been accessible at present energies until recently \cite{directphoton,jetsHP}. However, additional statistics and refinements in the experimental analysis, and also on the theoretical implementation --- in particular, the development of a Monte-Carlo approach to jet medium-modifications \cite{MC}, are required before physics conclusions can be drawn. For this reason they will not be considered here. 

The plan of the paper is as follows: In Section \ref{hydro} the hydrodynamical model of the medium will be briefly described (with full details provided in \cite{Hirano2001,HT2002}). In Section \ref{pqcd} the pQCD formalism used to compute single- and double-inclusive particle production will be outlined. In Section \ref{mixed} the implementation of energy loss will be reviewed. Finally, our results will be shown and discussed in Section \ref{results}, and the conclusions presented in Section \ref{conclu}.

\section{Hydrodynamical medium modeling} 
\label{hydro}

The energy loss of an energetic parton is determined by the properties of the medium along the parton 
trajectory~\cite{reports}. In relativistic heavy-ion collisions,
the medium itself evolves dynamically together with the
propagation of high-$p_{T}$ partons. Full three-dimensional
(3D) ideal hydrodynamics \cite{Hirano2001,HT2002} is employed
in our calculations to describe the space-time evolution of the hot matter created
in Au+Au collisions at top RHIC energy ($\sqrt{s_{NN}} = 200$ GeV).

We solve the equations of energy-momentum conservation
\begin{eqnarray}
\label{eq:hydro}
\partial_{\mu}T^{\mu\nu}=0
\end{eqnarray}
in full 3D space $(\tau,x,y,\eta_s)$ under the assumption that
local thermal equilibrium is reached at an initial
time $\tau_{0}$ =0.6 fm/$c$ and maintained thereafter until freeze-out. Here $\tau$, $\eta_s$,
$x$ and $y$ are proper time, space-time rapidity, and
two transverse coordinates perpendicular to the beam axis, respectively.
Ideal hydrodynamics is
characterized by the following energy-momentum tensor:
\begin{equation}
T^{\mu\nu}=(\epsilon+P)u^{\mu}u^{\nu}-Pg^{\mu\nu},
\end{equation}
where $\epsilon$, $P$ and $u^{\mu}$ are energy density, pressure and
local four velocity, respectively. We neglect the finite net-baryon
density which is small near mid-rapidity at RHIC where all calculations will be performed.
For the high temperature ($T>T_{c}=170$ MeV) Quark-Gluon Plasma (QGP) phase we use the equation of
state (EOS) of an ultra-relativistic massless parton gas ($u$, $d$, $s$ quarks and gluons) with a bag pressure $B$:
\begin{eqnarray}
p=\frac{1}{3}(\epsilon-4B).
\end{eqnarray}
The bag constant is tuned to be $B^{\frac{1}{4}}=247$\,MeV to
match the pressure of the QGP phase to that of a hadron resonance gas
at critical temperature $T_{c}=170$\,MeV.
On the other hand, a hadron resonance gas
model including all hadrons up to the mass of the $\Delta(1232)$
is employed for the low temperature phase ($T<T_{c}$).
The hadron resonance gas EOS employed in this study
implements chemical freeze-out
at $T_{\mathrm{ch}}=170$ MeV \cite{HT2002}
for simultaneous reproduction of particle ratios and spectra
at low $p_{T}$.
We assume that the hadronic matter behaves like a perfect fluid 
all the way down to the kinetic decoupling at $T^{\mathrm{dec}} = 100$ MeV.

The initial entropy distribution in the transverse plane is proportional
to a linear combination of the number density of participants, $\rho_{\mathrm{part}}$, 
and that of binary collisions, $\rho_{\mathrm{coll}}$~\cite{HHKLN2006} as given below:
\begin{equation}
s \propto (1-x) {\rho_{\rm part} } + x \rho_{\rm coll}.
\end{equation}
The proportionality constant and the fraction of soft and hard components 
(specified by $x$ = 0.15)
are so chosen that the centrality dependence of charged particle multiplicity
observed by PHOBOS \cite{PHOBOS_Nch} is reproduced~\cite{HHKLN2006}. 

For the initial condition of the flow velocity, Bjorken's scaling solution \cite{Bjorken},
namely $u_{x}(\tau_{0}) = u_{y} (\tau_{0}) 
= u_{\eta_{s}}(\tau_{0}) = 0$, is employed.
The hydrodynamic model has been tested against experimental data at RHIC energies such as centrality and $p_T$
dependencies of spectra for pions, kaons and protons up to 1.5 GeV/$c$ \cite{Hirano:2003pw,Hirano:2004en,HHKLN2006}.
We will use these hydrodynamic solutions, in particular the local energy density $\epsilon$ along the 
trajectory of partons, to fit various high-$p_T$ data and to reliably constrain
the transport coefficient $\hat{q}$. The hydrodynamic results such as temperature
and energy density as functions of the space-time position are publicly available \cite{hydro-site}. 

\section{Perturbative cross sections}
\label{pqcd}

The production of a hadron $h$ at transverse momentum $p_T$ and rapidity $y$ can be described by the factorized expression
\begin{equation}
\frac{d\sigma^{AA\to h}}{dp_Tdy}=\int \frac{dx_2}{x_2} \frac{dz}{z} \sum_{i,j}x_1f_i^A(x_1,Q^2)x_2f_j^A(x_2,Q^2)\frac{d\hat\sigma^{ij\to kl}}{d\hat t}D_{k\to h}(z,\mu_F^2),
\label{eq:crosssec}
\end{equation}
where $f_i^A(x,Q^2)$ are the parton distribution functions (PDFs) for a colliding object, either a proton 
or a nucleus $A$, $x$ is the momentum fraction of the hadron carried by the parton, and 
$D_{k\to h}(z,\mu_F^2)$ are the fragmentation functions for a parton $k$ into a hadron $h$, 
with $z$ the momentum fraction of the parton carried by the hadron.
We utilize CTEQ6L  and CTEQ6M \cite{Pumplin:2002vw} free proton parton densities depending on whether 
the calculation is at leading order (LO) or next-to-leading order (NLO).
We take the factorization and renormalization scales as $Q^2=(p_T/z)^2$ and the fragmentation scale as $\mu_F=p_T$ for light meson production, see e.g. \cite{Eskola:2002kv}. For the case of heavy quarks, we set all scales equal to
$Q^2=\mu_F^2=M^2+p_T^2$ for consistency with the comparison to experimental data in the proton-proton case \cite{Cacciari:2005rk}.
When the fragmentation takes place in a medium, the corresponding medium-modified fragmentation function is modeled as \cite{Wang:1996yh}
\begin{equation}
D^{\rm med}_{k\to h}(z,\mu_F^2)=\int\frac{d\zeta}{1-\zeta}P(\zeta)D^{\rm vac}_{k\to h}\left(\frac{z}{1-\zeta}, \mu_F^2\right),
\label{eq:mmff}
\end{equation}
where $P(\zeta)$ is the probability distribution of a fractional energy loss, $\zeta=\Delta E/E$, of the fast parton in the medium \cite{Baier:2001yt} and the vacuum fragmentation function, $D^{\rm vac}_{k\to h}$, is taken from a known set, in
this case from de Florian, Sassot and Stratmann \cite{deFlorian:2007aj}. Eq. (\ref{eq:mmff}) assumes that no modification of the virtuality evolution takes place in the medium, the only effect being the energy loss of the leading parton when traveling through the medium. This assumption has been studied recently \cite{Armesto:2007dt}, mainly motivated by the need of a more sophisticated approach for more exclusive observables.  However, the fact that the medium-induced gluon radiation does not present collinear or soft divergences, due to formation time effects, indicates that the corrections to (\ref{eq:mmff}) are ${\cal O}(1/p_T)$.

In the case of proton-proton collisions, the description of the data using Eq. (\ref{eq:crosssec}) is good even at lowest order (LO, with appropriate K-factors) for light hadrons produced at central rapidities \cite{Eskola:2002kv}. 

For particles containing heavy quarks, the Fixed-Order plus
Next-to-Leading-Log  (FONLL) approach \cite{Cacciari:1998it} is used to resum the potentially
large $\log(M_Q/p_T)$ terms, with $M_Q$ the mass of the heavy quark. This scheme
leads to a good description of the Tevatron data \cite{Cacciari:2002pa,Cacciari:2003uh,Cacciari:2003zu}. At RHIC, both the
PHENIX and the STAR Collaboration have measured the yield of non-photonic
electrons  \cite{Adare:2006hc,Abelev:2006db}, which are expected to be dominantly produced by the
semileptonic decay of the heavy quarks inside the hadrons. In  this case,
the comparison of data with FONLL is reasonably good for PHENIX, while the
STAR data appear underestimated by a factor  $\sim 5$. The discrepancy between
the two experimental data set should obviously be better understood before
addressing in more detail the agreement with the theoretical prediction.

The relative contribution of charm and beauty to the final yield of
electrons is not known precisely, as it suffers from large theoretical
uncertainties due to the variation of the actual value of the
mass of the quarks and unknown higher orders \cite{Cacciari:2005rk}, see \cite{Adare:2009ic} for recent
experimental efforts. This uncertainty is particularly relevant for the
medium effects as we will discuss below (also \cite{Armesto:2005mz}).

Besides single-particle inclusive cross sections, we are interested in the case of two-particle correlations. 
The corresponding expression is similar to Eq. (\ref{eq:crosssec}), but now two partons created in the hard 
scattering fragment independently. The fragmentation of each parton is again described by 
$D^{\rm vac}_{k\to h}$ or $D^{\rm med}_{k\to h}$ for the vacuum and the medium respectively. 
Notice that here we will be interested in two-particle correlations which are well separated in 
azimuthal angle (back-to-back correlations), so that correlations from the fragmentation of a single 
parton into two final hadrons are negligible~\cite{Ceccopieri:2007ip,Majumder:2004br}.

\section{Energy loss implementation}
\label{mixed}

The medium-modified fragmentation functions are computed using the Salgado-Wiedemann quenching weights $P(\Delta E/\omega_c,R)$ \cite{Salgado:2003gb}. They depend on two quantities, $\omega_c$ and $R$, which for a static medium of length $L$ and transport coefficient $\hat{q}$, are given by
\begin{equation}
\omega_c=\frac{1}{2}\hat{q} L^2,\ \ R=\omega_c L.
\end{equation}
In a dynamical medium we make use of the following scaling relations \cite{Salgado:2002cd,Dainese:2004te,Renk:2006pk,Renk:2006sx,Bass:2008rv}
\bea
\omega_c^{eff}(x_0,y_0,\tau_{\rm prod},\phi)=\int d\xi \, \xi \hat{q}(\xi),\label{eq:eff1}\\ 
\protect[ \hat{q}L ]^{eff}(x_0,y_0,\tau_{\rm prod},\phi)=\int d\xi \, \hat{q}(\xi),\label{eq:eff2} \\
R^{eff}(x_0,y_0,\tau_{prod},\phi)=\frac{2 [\omega_c^{eff}(x_0,y_0,\tau_{\rm prod},\phi)]^2}{(\hat{q}L)^{eff}(x_0,y_0,\tau_{\rm prod},\phi)}\ , \label{eq:eff3}
\eea
corresponding to an equivalent static scenario to write the effective quantities. 
These quantities are for a parton produced at mid-rapidity where the measurements under consideration 
have been performed, and at transverse position $(x_0,y_0)$ and time $\tau_{\rm prod}$. 
Each parton traverses the medium with a straight-line trajectory given by azimuthal angle $\phi$ and
with local transport coefficient $\hat{q}(\xi)$.  As such, the problem is reduced to specifying the relation 
between the local value of the transport coefficient at a given point, $\xi$, on the trajectory and the 
medium properties as given by our hydrodynamical model. Based on the estimates in~\cite{Baier:2002tc}, we define 
\begin{equation}
\hat{q}(\xi)=K\hat q_{\rm QGP}(\xi)\simeq K\cdot 2  \epsilon^{3/4}(\xi). 
\label{eq:qhateps}
\end{equation}
The energy density along the parton trajectory $\epsilon(\xi)$ is then taken from the hydrodynamical model of the medium and $K$   is a fit parameter when comparing to experimental data. This parameter then
can indicate any departure with respect to the ideal gas $\hat q_{\rm QGP}$ estimated in \cite{Baier:2002tc}. 

The production point of the hard scattering is characterized by a production weight $w(x_0,y_0)$, as computed 
through the product of the profile functions of the two nuclei,
\begin{equation}
w(x_0,y_0)=T_{Au}(x_0,y_0)T_{Au}(\vec{b}-(x_0,y_0)),
\end{equation}
with the profile function given by a 3-parameter Fermi distribution  at a
 given impact parameter $\vec{b}$ taken from Ref. \cite{De Jager:1987qc}.
The trajectory is then defined by an angle $\phi$ and the average of a given quantity $\cal O$ is computed as
\begin{equation}
\langle {\cal O}\rangle=\frac{1}{N}\int d\phi dx_0dy_0 w(x_0,y_0){\cal O}(x_0,y_0,\phi),
\label{ave}
\end{equation}
with $N=2\pi\int dx_0dy_0 w(x_0,y_0)$. For example, we may compute the average value of $\hat{q}$ as follows:
\begin{equation}
\langle \hat{q}\rangle=\frac{1}{N}\int d\phi dx_0dy_0 w(x_0,y_0)\frac{\left[[\hat{q}L]^{eff}(x_0,y_0,\phi)\right]^2}{2 \omega_{c}^{eff}(x_0,y_0,\phi)}\,.
\label{eq:qhataverage}
\end{equation}
The medium-modified fragmentation functions are computed in the same manner:
\begin{equation}
\left\langle D^{\rm med}_{k\to h}(z,\mu^2_F)\right\rangle=\frac{1}{N}\int d\phi dx_0dy_0 w(x_0,y_0)\int \frac{d\zeta}{1-\zeta}P(x_0,y_0,\phi,\zeta)D^{\rm vac}_{k\to h}\left(\frac{z}{1-\zeta},\mu^2_F\right),
\end{equation}
where $P(x_0,y_0,\phi,\epsilon)$ is the corresponding quenching weight for the effective parameters (\ref{eq:eff1}) and (\ref{eq:eff3}). In the case of two-particle correlations, two fragmentation functions are computed for particles traveling in opposite directions -- i.e. measuring different medium lengths and density profiles:
\begin{equation}
D^{\rm med}_{k\to h}(x_0,y_0,\phi,z,\mu^2_F)=\int \frac{d \zeta}{1-\zeta}P(x_0,y_0,\phi, \zeta)D^{\rm vac}_{k\to h}\left(\frac{z}{1-\zeta},\mu^2_F\right),
\end{equation}
with the average given by
\begin{eqnarray}
&& \left\langle D^{\rm med}_{k\to h_1}(z_1,\mu^2_{F,1})D^{\rm med}_{l\to h_2}(z_2,\mu^2_{F,2})\right\rangle=\\ 
&& \frac{1}{N}\int d\phi dx_0dy_0 w(x_0,y_0) D^{\rm med}_{k\to h_1}(x_0,y_0,\phi,z_1,\mu^2_{F,1})D^{\rm med}_{l\to h_2}(x_0,y_0,\pi-\phi,z_2,\mu^2_{F,2}).
\end{eqnarray}

\subsection{Energy loss for times prior to hydrodynamic behavior}

In a dynamical medium like the one considered here, there is an ambiguity on the value of the transport coefficient, defined by Eq. (\ref{eq:qhateps}), for values smaller than the thermalization time $\tau_0$. One extreme case is to take $\hat q(\xi)=0$ for $\xi<\tau_0$. The absence of any energy-loss effect for these very early times is a strong assumption 
since thermalization is not necessary in the approach in which the quenching weights have been computed~\cite{reports}. 
To quantify this uncertainty, we consider three different extrapolations for the time from the hard parton production
to the thermalization time ($\tau_0=0.6$ fm/$c$ in the present calculation):
\begin{itemize}
\item Case i): $\hat q(\xi)=0$ for $\xi<\tau_0$;
\item Case ii): $\hat q(\xi)=\hat q(\tau_0)$ for $\xi<\tau_0$; and
\item Case iii): $\hat q(\xi)=\hat q(\tau_0)/\xi^{3/4}$ for $\xi<\tau_0$.
\end{itemize}
These extrapolations range from the most extreme assumption of no effect at all before the thermalization time (Case i)) 
to a continuous interaction from the production time (taken to be $\tau_{\rm prod}\simeq 0.04$ fm/$c$) with a 
free-streaming medium with energy density dropping as $\epsilon(\xi)\sim1/\xi$ (Case iii)).
As it turns out, the extracted value of the parameter $K$ in Eq. (\ref{eq:qhateps}) is only modestly sensitive to
different assumptions about the very early time.  

Both the quenching weights and the hydrodynamical profiles are taken 
from the publicly available codes \cite{qw-site} and \cite{hydro-site}, respectively.

\section{Results and discussion}
\label{results}

The experimental data used in our analysis are given in terms of the nuclear modification factors
\begin{eqnarray}
R_{AA}=\frac{dN_{AA}/dp_T^2dy}{\left< N_{\rm coll}\right> dN_{pp}/dp_T^2dy},&& I_{AA}=\frac{D_{AA}(z_T,p_T^{\rm trig})}{D_{pp}(z_T,p_T^{\rm trig})}
\label{eq:raa}
\end{eqnarray}
for single- and double-inclusive measurements, respectively. They are computed using the corresponding 
medium-modified fragmentation functions, (\ref{eq:mmff}), in the cross sections, (\ref{eq:crosssec}), 
for the $AA$ case and the corresponding unmodified (vacuum) ones for the proton-proton case appearing 
in the denominators of (\ref{eq:raa}). In (\ref{eq:raa}), the hadron-triggered fragmentation function 
is defined as \cite{Wang:2003mm}
\begin{equation}
D_{AA}(z_T,p_T^{\rm trig})\equiv p_T^{\rm trig}\frac{d\sigma_{AA}^{h_1h_2}/dy^{\rm trig}dp_T^{\rm trig}dy^{\rm assoc}dp_T^{\rm assoc}}{d\sigma^{h_1}_{AA}/dy^{\rm trig}dp_T^{\rm trig}}\,.
\label{eq:daa}
\end{equation}
Here, $z_T=p_T^{\rm assoc}/p_T^{\rm trig}$, and we use as factorization scales the $p_T$ of the hadrons. 
We highlight that the quantities defined in this manner are, in general, very different from the fragmentation 
functions measured, for example, in $e^+e^-$ annihilation at similar virtualities due to the strong bias 
produced by triggering on a steeply falling perturbative particle spectrum.

The data sets included in our analysis are all for Au+Au collisions at 
top RHIC energy, $\sqrt{s_{NN}} = 200$ GeV and are listed below.
\begin{enumerate}
\item 'Light': PHENIX data on $R_{AA}$ for $\pi^0$ with $p_T^{\pi^0}> 5$ GeV in the $0-5$ \% centrality class  \cite{phenix}, 14 data points.
\item 'Dihadron': STAR data on $I_{AA}$ for Au+Au/d+Au for charged particles in the $0-5$ \% centrality class \cite{Adams:2006yt}, for $p_T^{\rm trig}=8-15$ GeV and $p_T^{\rm trig} > p_T^{\rm assoc} > 3.0$ GeV, 6 data points\footnote{For computational time reasons, we make the calculations with fixed $p_T^{\rm trig}=10$ GeV. We have checked that the results are not sensitive to changes in this value. With this choice of $p_T^{\rm trig}$ and $p_T^{\rm assoc}$, the $D_{AA}(z_T,p_T^{\rm trig})$ measured \cite{Adams:2006yt} in d+Au collisions at top RHIC energy is well described.}.
\item 'Heavy': $R_{AA}$ for non-photonic electrons with $p_T^e>3$ GeV, for $0-10$ \% from PHENIX \cite{Adare:2006nq}, 11 data points; and for $0-5$ \% from STAR \cite{Abelev:2006db}, 9 data points.
\end{enumerate}

In order to obtain the best value of the $K$-parameter (or equivalently $\hat q$) we perform a modified 
$\chi^2$-analysis that treats statistical and systematic uncertainties in a consistent
manner, as detailed in~\cite{phenix}. For any set of data with central values $\{y_i\}$, three types of 
uncertainties $\sigma_i^t$, $t=A$ (statistical plus uncorrelated systematics added in quadrature), $B$ (correlated systematics) and $C$ (normalization uncertainties in \%), are considered and 
a modified  $\tilde \chi^2$ for each set of theoretical values $\{y_i^{\rm th}\}$ 
(characterized by one given value of parameter $K$ in the present case) is obtained by minimizing
\begin{eqnarray}
\tilde\chi^2&=&\left[ \  \sum_{\rm exp.\  points}  \frac{[y_i+\varepsilon_B\sigma_i^B+\varepsilon_C y_i \sigma^C-y^{th}_i]^2}{\tilde{\sigma_i}^2}\right]  +\varepsilon_B^2+\varepsilon_C^2,\\
\tilde{\sigma_i}&=&\sigma_i^A\left(\frac{y_i+\varepsilon_B\sigma_i^B+\varepsilon_C y_i \sigma^C}{y_i}  \right),
\end{eqnarray}
with respect to the two parameters $\varepsilon_B$, $\varepsilon_C$.   Note that our calculations currently
do not explicitly include theoretical systematic uncertainties in the above formulation.  
Instead we consider various modifications of the theoretical calculations and their impact of the experimentally
constrained results.  In the case of the the 'heavy' set,
there is a theoretical uncertainty of $\pm 8$ \% coming from the variation of the renormalization and 
factorization scales and the heavy quark mass, see  \cite{Armesto:2005mz}. 
Since as will be shown, the theory tends to predict less suppression than the experimental data, we 
take the lowest values inside this band as the theoretical input $\{y_i^{\rm th}\}$ for the fits.

It is notable that in the 'light' case, we follow the treatment of systematic uncertainties in ~\cite{phenix}
where the type B uncertainties allow a tilt of the data points (considering cases of correlation or
anti-correlation as a function of $p_T$).  In the 'dihadron' case, we include the statistical uncertainties
quoted in ~\cite{Adams:2006yt} and an additional 7\% normalization uncertainty as used in 
the constraint analysis~\cite{Nagle:2008fw}.  In the 'heavy' case, both experiments quote systematic uncertainties
which are included in the analysis here.  Note that we consider the type B uncertainties to move the data points
up and down together (and do not consider the anti-correlation case here).  
We discuss these uncertainties for the 'heavy' case in more detail later.

We present the results for different values of the proportionality factor between the transport coefficient and the energy density, Eq. (\ref{eq:qhateps}), $K=K^\prime/1.46$, where  $K^\prime=0.5, 1, 2, 3,\dots,$ $20$.  For the 'heavy' case
we extended the range up to $K^\prime$= 40.  This grid in $K$ is then used to obtain the best $\tilde \chi^2$ values. 

\subsection{Light hadrons}

Our first study corresponds to a global fit of $R_{AA}$ and $I_{AA}$ within the model and assuming a frozen value of $\hat q$ for the very early times ---Case ii) previously discussed.  Additionally, the EKS98 set \cite{Eskola:1998df} 
is taken for the nuclear corrections to the PDFs. This fixes our canonical choices for the calculation.

\begin{figure}
\begin{center}
\includegraphics[width=0.7\textwidth]{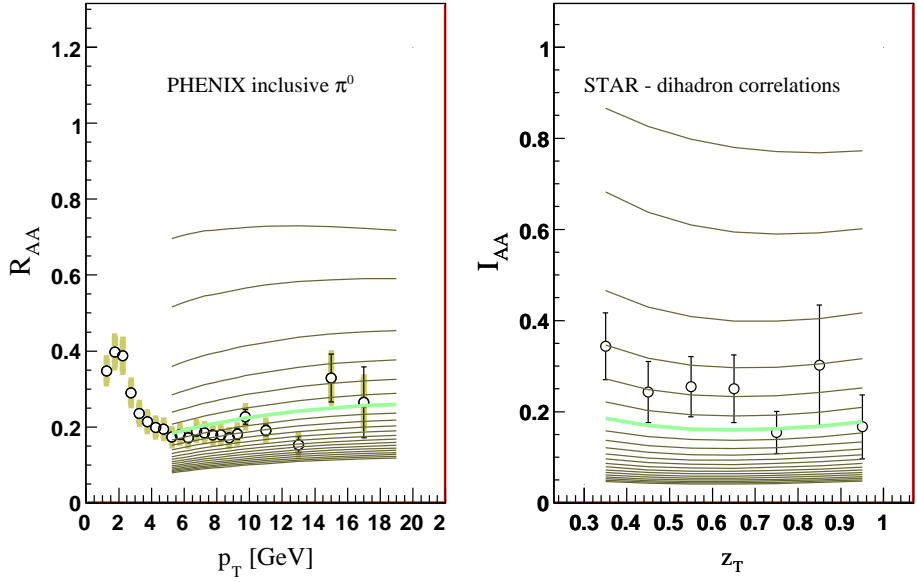}
\end{center}
\caption{Nuclear modification factors $R_{AA}$ for single-inclusive (left) and $I_{AA}$ for hadron-triggered fragmentation functions (right) for different values of $K=K^\prime/1.46$, where  $K^\prime=0.5, 1, 2, 3,\dots,$ $20$.   The theoretical curves are for the canonical calculation (i.e. EKS98 for the nPDF and Case ii) for the treatment of very early times).  Shown with a green line is the curve corresponding to the minimum of the common fit to $R_{AA}$ and $I_{AA}$ data: $K=4.1$.}
\label{fig:raa1}
\end{figure}

In Fig. \ref{fig:raa1} we plot the curves for different $K$-factors for both $R_{AA}$ and $I_{AA}$ and compare 
with the neutral pion data suppression from PHENIX and the corresponding suppression of the back-to-back 
hadron-triggered fragmentation functions from STAR.   Shown in Fig. \ref{fig:chiK} as solid lines are the
calculated $\tilde \chi^2$ results for both $R_{AA}$ and $I_{AA}$ comparisons.  For our canonical choice, the
constraint on $K$ agrees within one standard deviation uncertainties.  We then sum the two $\tilde \chi^2$ values
as a function of $K$ and obtain a common global fit result of $K= 4.1 \pm 0.6$.

\subsubsection{Comparison of different prescriptions for $\tau<\tau_0$}

\begin{figure}
\begin{minipage}{0.49\textwidth}
\begin{center}
\includegraphics[width=0.85\textwidth]{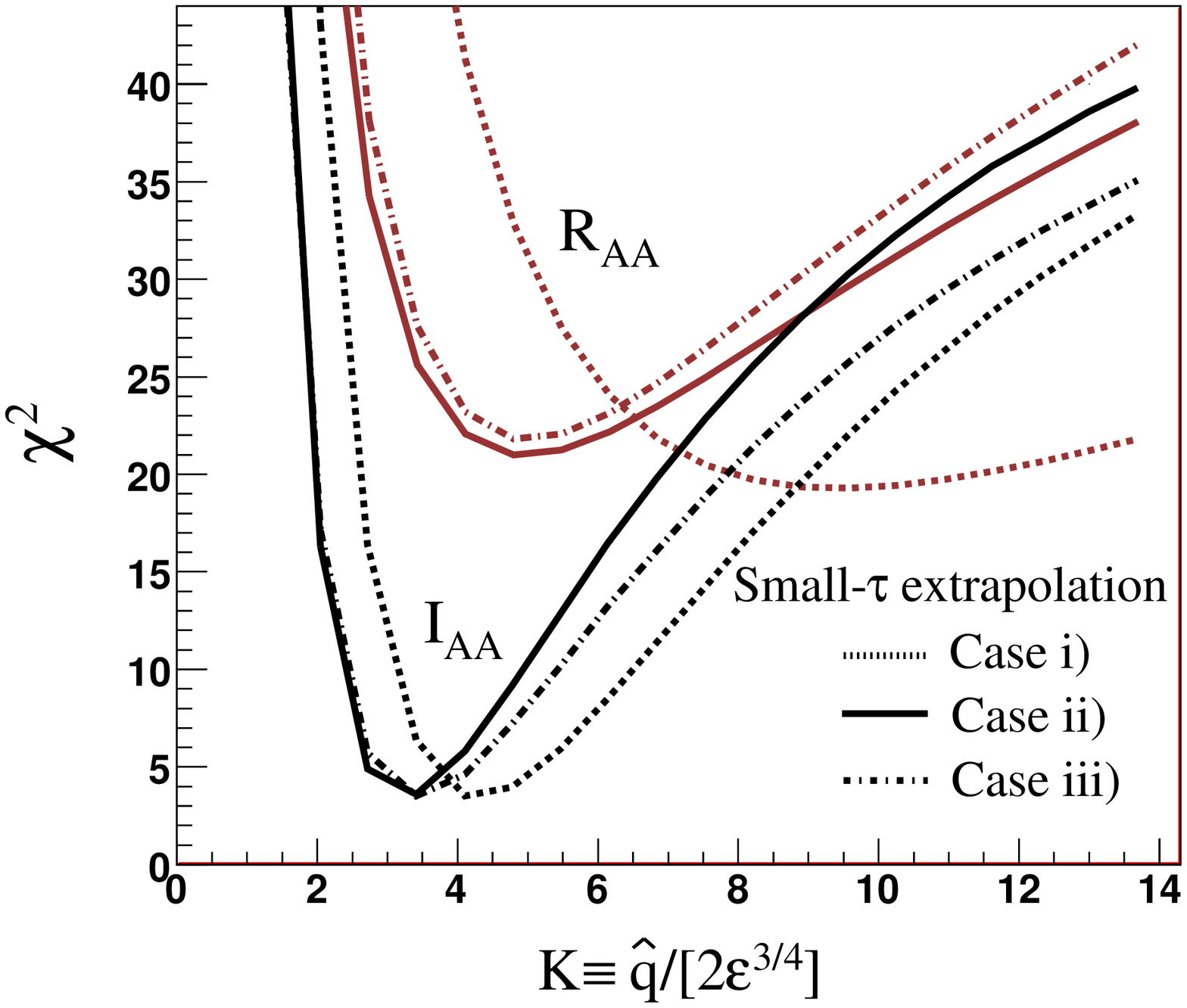}
\end{center}
\end{minipage}
\hfill
\begin{minipage}{0.51\textwidth}
\begin{center}
\includegraphics[width=\textwidth]{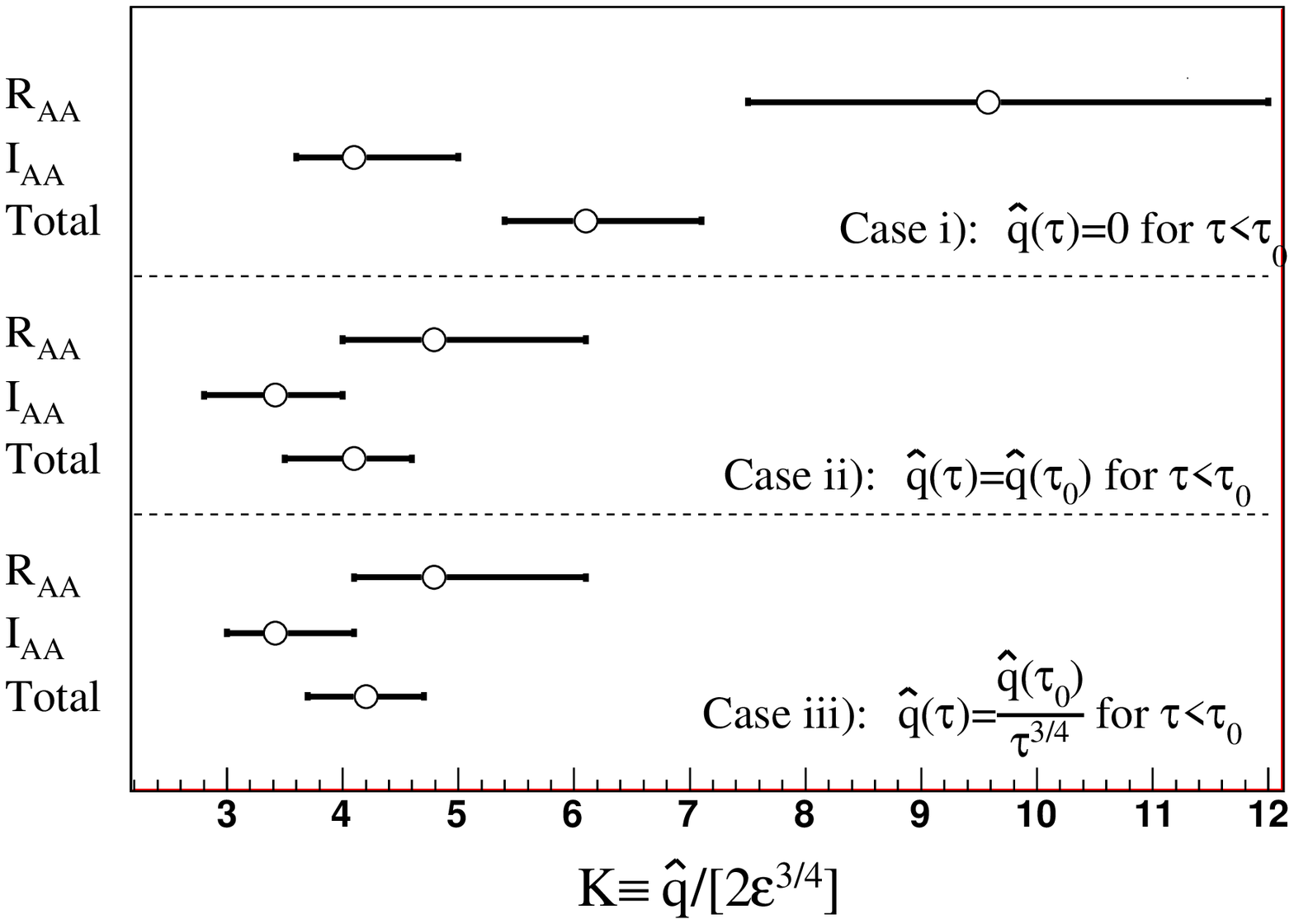}
\end{center}
\end{minipage}
\caption{Left: $\tilde \chi^2$-values for different values of $K$ for light hadrons and for the three different extrapolations for $\xi<\tau_0$ considered here: Case i), Case ii) and Case iii) are plotted with dotted, solid and dotted-dashed lines respectively. Red lines correspond to single-inclusive $\pi^0$ data from PHENIX ($R_{AA}$) and black ones to the double-inclusive measurements by STAR ($I_{AA}$). Right: the corresponding central values (minima of the $\tilde \chi^2$) and the uncertainties computed by considering  $\Delta \tilde \chi^2=1$ are shown.}
\label{fig:chiK}
\end{figure}

We now repeat the analysis for the two other assumptions for $\hat q$  for times $\tau<\tau_0=0.6$ fm/$c$. The result of this analysis leads to the computed values of the $\tilde \chi^2$ for light hadrons $R_{AA}$ and $I_{AA}$ shown in Fig. \ref{fig:chiK} (dotted and dotted-dashed lines) and to the corresponding determination of the central values and the uncertainty band for $\Delta \tilde \chi^2=1$.  The results are relatively insensitive to the choice of very early time treatment,
except for $R_{AA}$ in Case i).  This can be understood by the following example contribution: the parton travels 
for the first $0.6$ fm/$c$ without any effect and the strong dependence of jet quenching on the in-medium path 
length forces the value of $K$ (and $\hat q$) to increase.

It is notable that in the way the averaged transport coefficient is computed in our approach, Eq. (\ref{eq:qhataverage}), 
the actual value of $\hat q$ is different for a given value of $K$ depending on the assumption for the initial times in the evolution --- this is simply because the average value of the energy density is different in the three cases. For example, the values at the minimum $K=4.1$ are $\langle\hat q\rangle \simeq $ 6.2, 8.7,  138.5 GeV$^2$/fm for the Cases i), ii) and iii) respectively. This also shows the large ambiguity in the determination of an average value of $\hat q$ for an evolving medium: unrealistically large values of $\langle \hat q\rangle$ are possible due to the averaging procedure, in particular when the extrapolation is divergent, as in Case iii). This is the reason why in this paper, we have made the choice of quoting only values of the $K$-factor instead of actual values of the averaged transport coefficients. The fact that our results are very solid under the two assumptions ii) and iii) provides strong support to the way in which the hydrodynamical medium and the jet quenching calculations are merged.

\subsubsection{The effect of the nuclear modification of parton densities}

We now explore the uncertainty associated with the choice of nuclear parton distribution functions. 
We consider three situations: i) nPDFs as given by the EKS98 analysis~\cite{Eskola:1998df}; ii) nPDFs 
as given by the EPS08 analysis~\cite{Eskola:2008ca}; iii) proton PDFs, i.e. without any nuclear modification. These 
three cases comprise two extreme situations\footnote{Notice that for a more formal analysis, the use 
of the uncertainties computed using the Hessian method as done in the recent EPS09 set~\cite{Eskola:2009uj}  
may be used.}.  
In this study, the impact parameter dependence of the nuclear PDF corrections is not taken into 
account explicitly.   This may impact the surface bias from parton energy loss, and is in part, why
we also consider the no shadowing case as an extreme.

\begin{figure}
\begin{minipage}{0.49\textwidth}
\begin{center}
\includegraphics[width=0.85\textwidth]{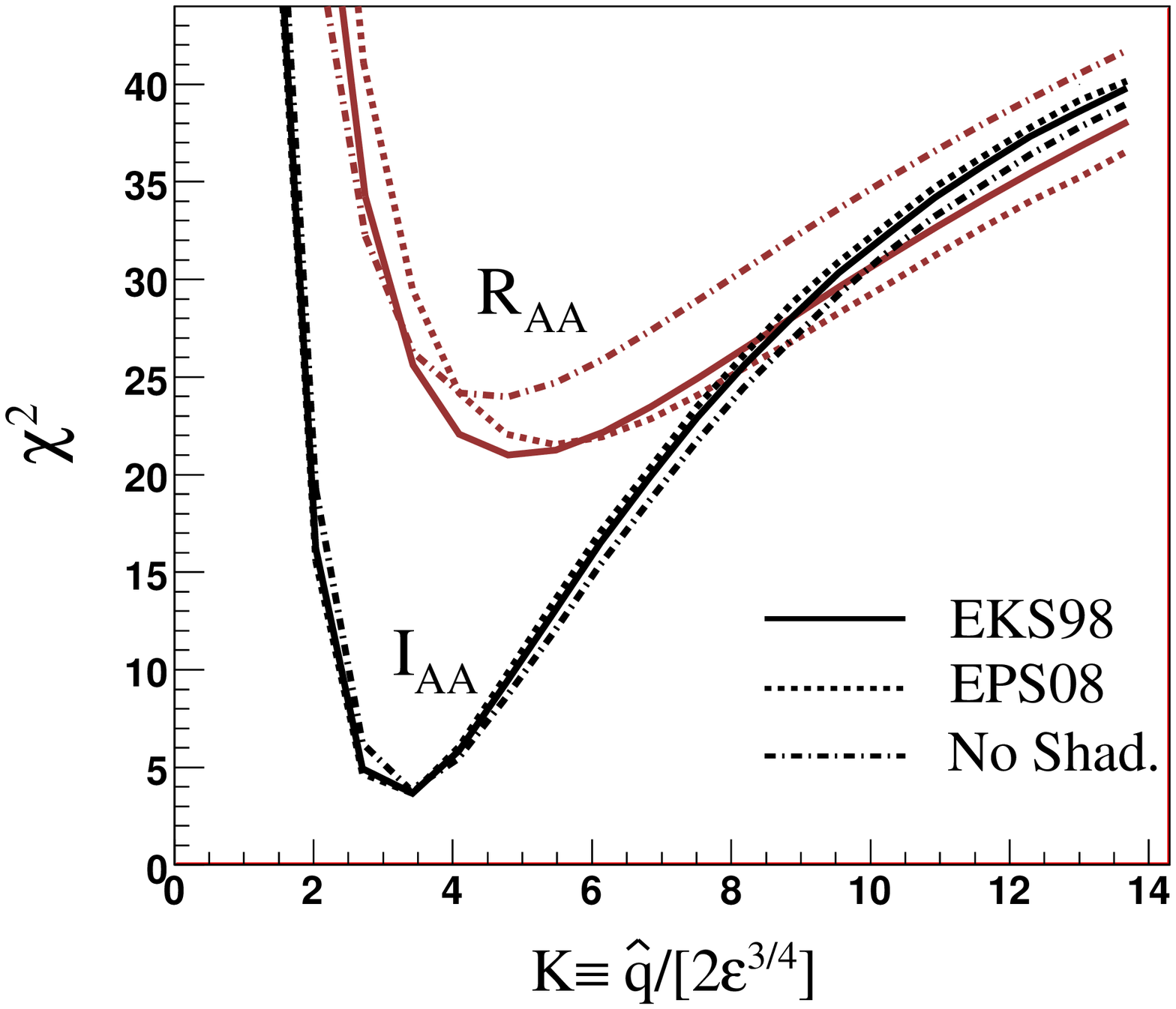}
\end{center}
\end{minipage}
\hfill
\begin{minipage}{0.51\textwidth}
\begin{center}
\includegraphics[width=\textwidth]{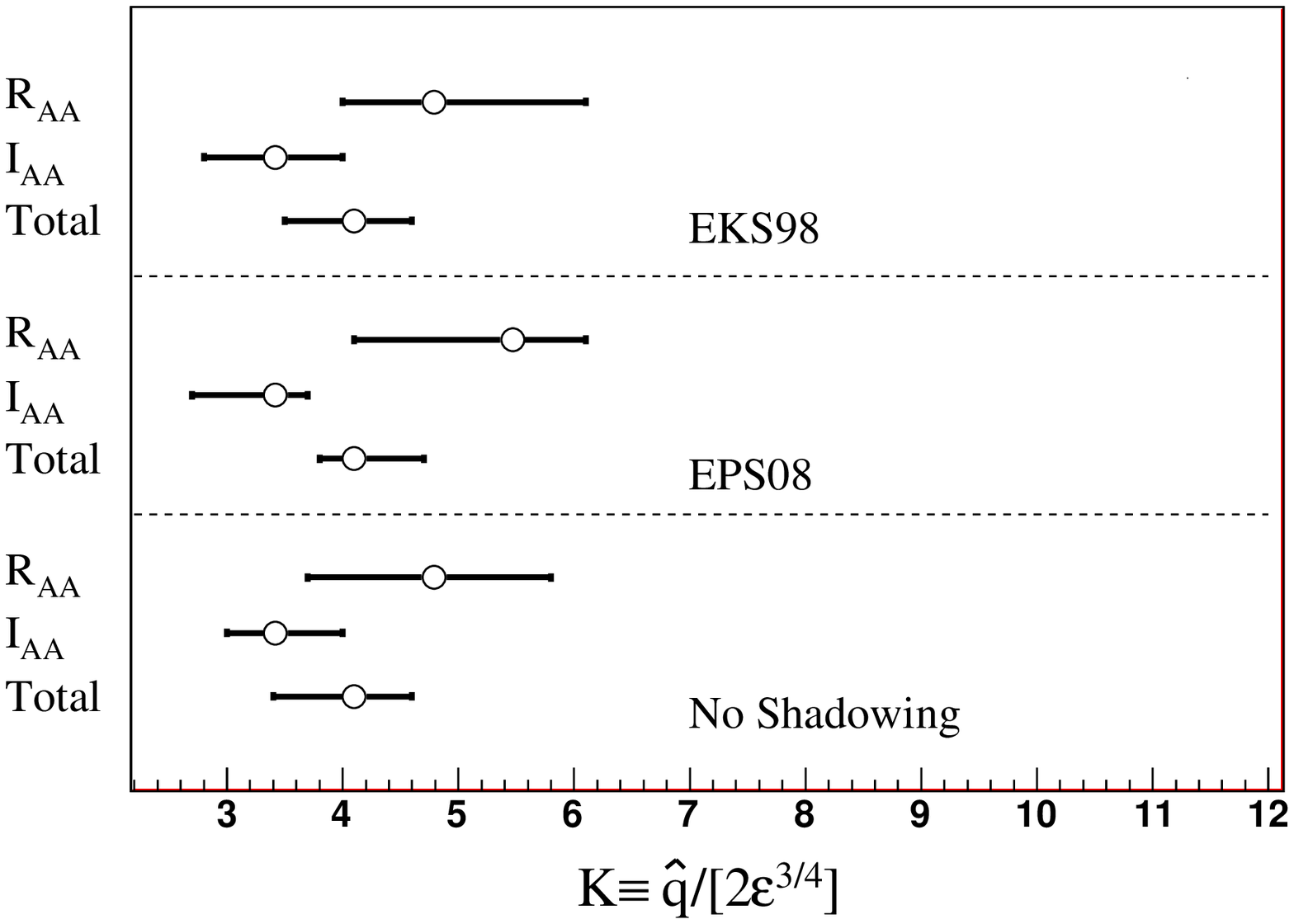}
\end{center}
\end{minipage}
\caption{Left: $\tilde \chi^2$-values for different values of $K$ for light hadrons and for three assumptions on the nuclear effects in PDFs: EKS98 (solid lines) \cite{Eskola:1998df}, EPS08 \cite{Eskola:2008ca} (dotted lines) and no shadowing corrections (dashed-dotted lines). As before, red lines correspond to single-inclusive $\pi^0$ data from PHENIX ($R_{AA}$) and black ones to the double-inclusive measurements by STAR ($I_{AA}$). Right: the corresponding central values (minima of the $\tilde \chi^2$) and the uncertainties computed by considering $\Delta \tilde \chi^2=1$ are shown.}
\label{fig:chiPDF}
\end{figure}

In Fig. \ref{fig:chiPDF} we plot again the value of $\tilde \chi^2$ for each choice of the nuclear PDFs and the corresponding central values and uncertainty bands at the $\Delta \tilde \chi^2=1$ level. The differences are, in this case, smaller than the uncertainties from the modeling of the initial times of medium evolution. Interestingly, the results for $I_{AA}$ are again somewhat less sensitive to these uncertainties than the ones for $R_{AA}$ --- the reason being the cancellation in the double ratio for the former. 

\subsection{Non-photonic electrons and heavy quarks suppression}

We now study the 'heavy' quark case of non-photonic electron data within our canonical set-up, 
i.e. Case ii) for extrapolations at small times and EKS98 for nuclear PDFs. 
Notice that within the radiative energy loss formalism, the energy loss of heavy quarks receives corrections due to the mass terms in the propagators whose net effect is a reduction of the amount of radiated energy. It is a generic prediction of the formalism that the energy loss of a heavy quark is smaller than that of a light quark. The computation of this feature, first pointed-out in Ref. \cite{Dokshitzer:2001zm}, is performed in the approximation $M/E\ll 1$ in Refs. \cite{Zhang:2003wk,Djordjevic:2003zk,Armesto:2003jh}. The quenching weights for the multiple-soft scattering approximation 
used here are found in \cite{qw-site}.

It was previously mentioned that the 
calculation has been performed in the $M/E\ll 1$ limit, when $M$ and $E$ are the mass and the energy of the 
heavy quark. This limit is reasonable for the charm quark case in the region under study. The case of the 
bottom quark is, however, less clear, as its mass and energy are comparable for a non-negligible region of 
the phase space under study. Further mass term corrections have not as yet been computed in this formalism. 

\begin{figure}
\begin{center}
\includegraphics[width=0.7\textwidth]{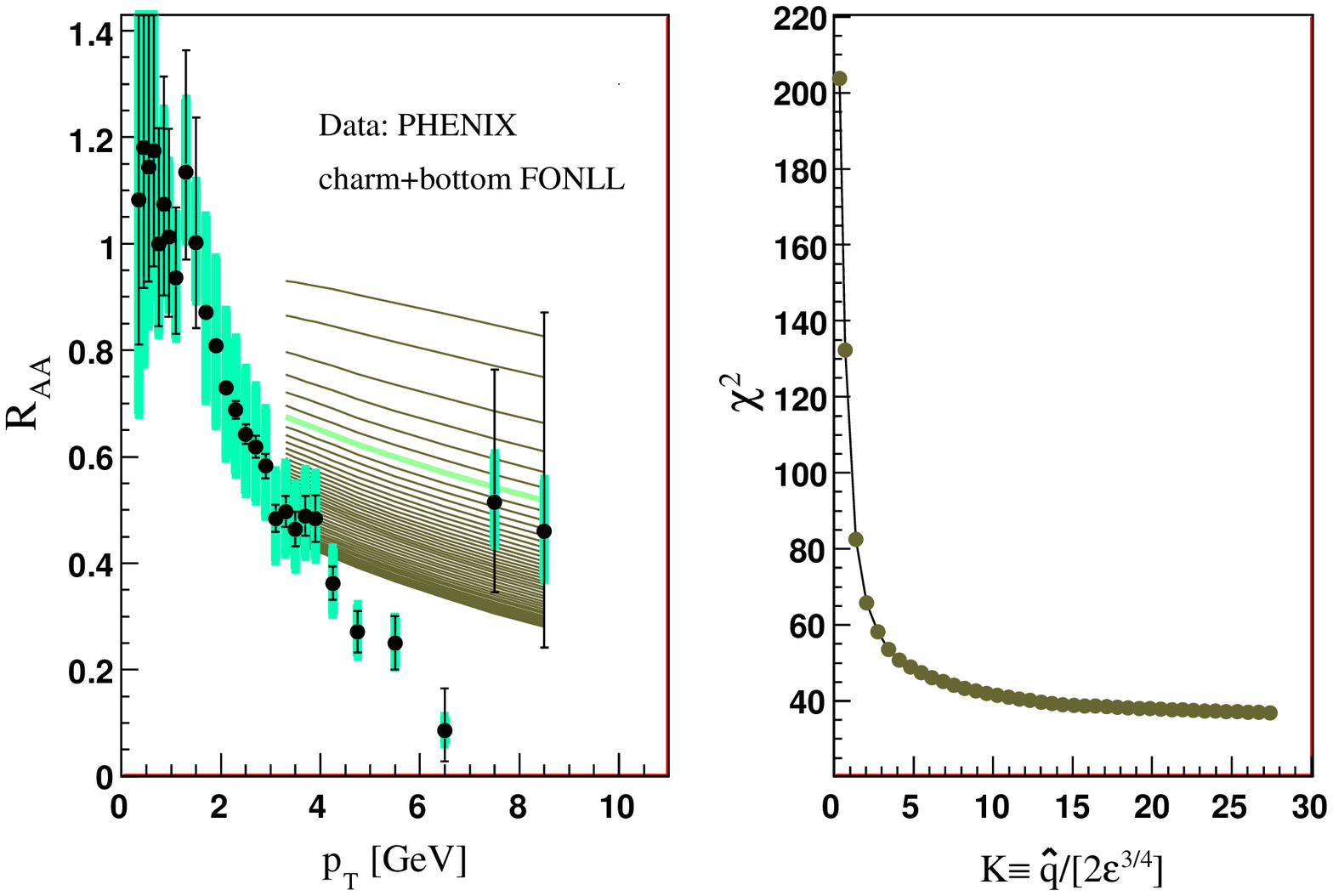}
\includegraphics[width=0.7\textwidth]{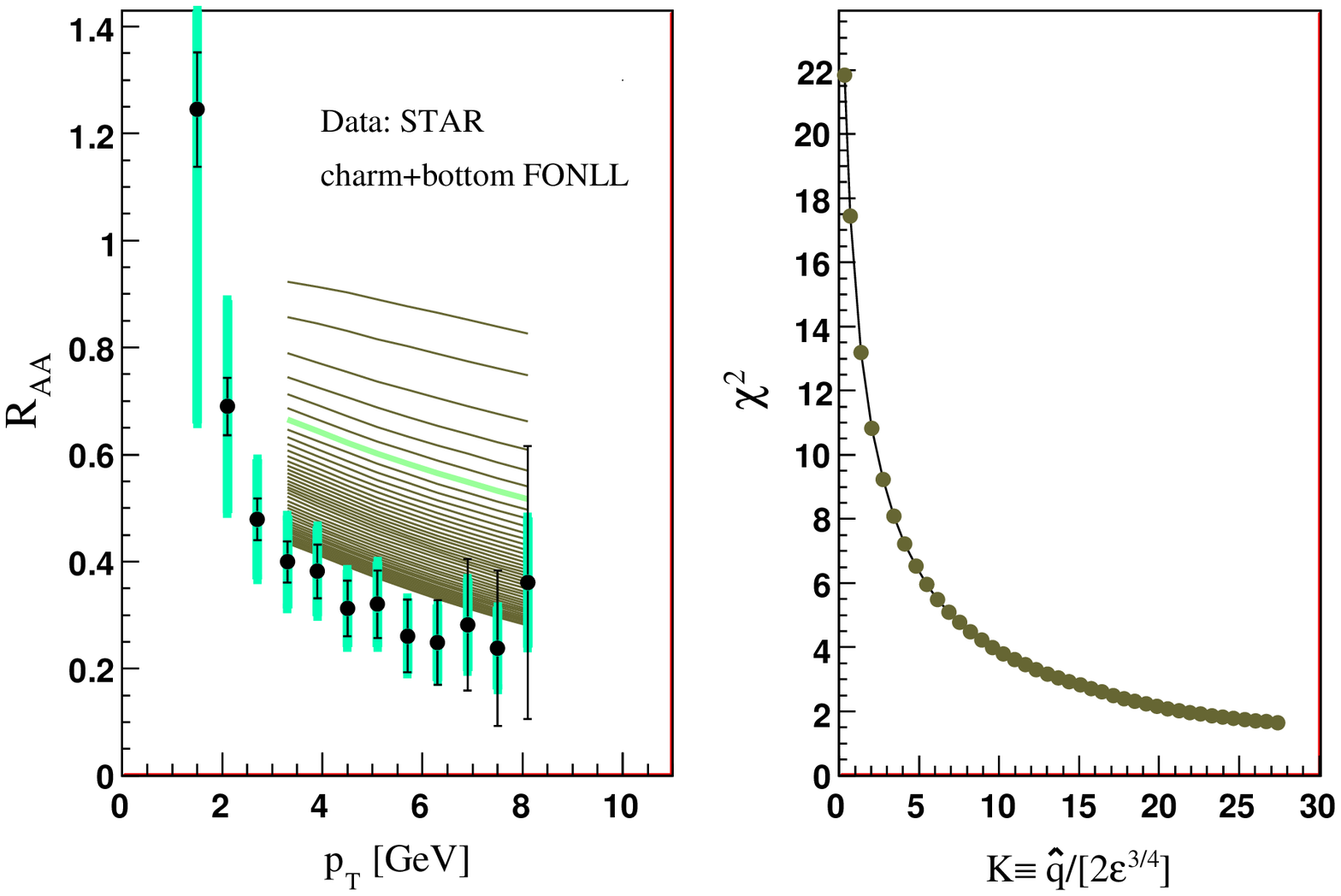}
\end{center}
\caption{Nuclear modification factor $R_{AA}$ and the computed values of $\tilde\chi^2$ for non-photonic electrons from heavy quark decays for the same values of the parameter $K$ as in Fig. \protect\ref{fig:raa1} --- the green line corresponds to the same value of $K$ as the green line if Fig. \protect\ref{fig:raa1}. The upper panels show the comparison with PHENIX data~\cite{Adare:2006nq} and the lower panels the comparison to the STAR data~\cite{Abelev:2006db}. The small difference in centrality, 0-10\% in PHENIX and 0-5\% in STAR is taken into account. The FONLL scheme is used in both calculations.}
\label{fig:heavy1}
\end{figure}

We now proceed as in the case of the light quarks and compute, for different values of the $K$-factor, the $R_{AA}$ of the non-photonic electrons from heavy-meson decays computed within the FONLL approach \cite{Cacciari:1998it}. As expected, the suppression is reduced due to the finite mass of the heavy quarks, which is in qualitative contradiction with the experimental data as shown in Fig.~\ref{fig:heavy1}.  As noted in Ref. \cite{Armesto:2005mz}, the unknown relative contribution from charm and bottom decays to the final electrons introduces an additional uncertainty in this comparison. Here, an $8\%$ theoretical uncertainty is used, as obtained by varying the heavy quark masses and the renormalization and factorization scales in the calculation \cite{Armesto:2005mz}.  The theoretical uncertainty band is likely larger, especially for larger values of $\hat q$. This uncertainty is not taken for the case in which only the charm quark contributes to the final electron yield as the uncertainty band is much smaller in this case.

We compute again the $\tilde \chi^2$ values for different values of the $K$-factor. In Fig. \ref{fig:heavy1} we do that for the central theoretical value of $R_{AA}$ diminished by the $8\%$ theoretical uncertainty as stated above. The first observation is that the values of $\tilde \chi^2$ never reach a minimum in the studied region of $K$ (thus favoring even larger values beyond the current calculation range).  Thus, the 'heavy' data sets appear incompatible with
the 'light' and 'dihadron' data sets within this theoretical framework.  However, it is striking that the minimum (within our $K=$factor range) is $\tilde \chi^{2}_{\rm min} \approx 37$ for the PHENIX data (with 11 data points) and 
$\tilde \chi^{2}_{\rm min} \approx 1.9$ for the STAR data (with 9 data points).  

For the STAR data sample~\cite{Abelev:2006db}, around $p_{T} \approx 5$ GeV, the typical statistical uncertainties 
are of order 32\%, the box type systematic uncertainties drawn on the points are of order 29\%, and the 
normalization uncertainty is 26\%.    In the publication, the
box type uncertainties are stated as uncorrelated~\cite{Abelev:2006db}.  If they are point-to-point uncorrelated, 
we should add them in quadrature with the statistical uncertainties (type A).  As an exercise, we use the 
full statistical constraint procedure with the theory comparison as a simple second-order polynomial 
($R_{AA} = a + b \times p_{T} + c \times p_{T}^{2}$).  One finds a best fit with 
$a=+0.70$, $b=-0.13$ GeV$^{-1}$, and $c=+0.01$ GeV$^{-2}$ with a total $\tilde \chi^{2} = 0.5$.  This 
represents an extreme in goodness-of-fit with a p-value = 99.8\%.  
In this paper, we have thus treated the STAR uncertainties
drawn as boxes around the points as standard type B (assuming the points are in fact correlated).  
Even in this case, the total $\tilde \chi^{2} = 1.1$ remains
an extremely good match (i.e. reflecting much smaller-than-expected fluctuations of the individual data points).  This results in
a rather improbable situation in that even a value $K = 4.1$ as determined by the 'light' and 'dihadron' data has a
reasonable p-value for agreement with the 'heavy' data set discussed here.  Note that a $\tilde \chi^2$ test is more
stringent in that in order for $K = 4.1$ to be the true answer, it means that $K = 30$ (for example) is not.  Thus,
since the $\tilde \chi^2$ for $K = 30$ is so small (and much smaller than for $K = 4.1$), it rules out
$K = 4.1$ by the $\Delta \tilde \chi^2$ test, but not by a simple p-value check.  

For the PHENIX data sample~\cite{Adare:2006nq}, around $p_{T} \approx 5$ GeV, the typical statistical uncertainties are of order 20\%, the type B uncertainties of
order 19\%, and the normalization uncertainty is 7\%.  In this analysis, we consider only the correlated case
for the type B uncertainties.  We again compare the data with a simple second-order polynomial and find the
best values as 
$a=+1.6$, $b=-0.4$ GeV$^{-1}$, and $c=+0.03$ GeV$^{-2}$ 
and a total $\tilde \chi^{2} = 15.2$.  This is considerably better than the best fit
with the parton energy loss calculation, but still only has a p-value $\approx$ 5\%.  We note that of order 30\% of the
total $\tilde \chi^{2}$ comes from the single data point at $p_{T} = 7$ GeV.  
The details discussed above regarding the PHENIX and STAR experimental uncertainties may be indicative of not yet fully understood systematic uncertainties and/or correlations amongst these uncertainties.
Future data sets, including
with upgraded detectors to specifically tag the displaced vertex from the D and B meson decays, will be crucial
to resolving these experimental issues.

\begin{figure}
\begin{center}
\includegraphics[width=0.7\textwidth]{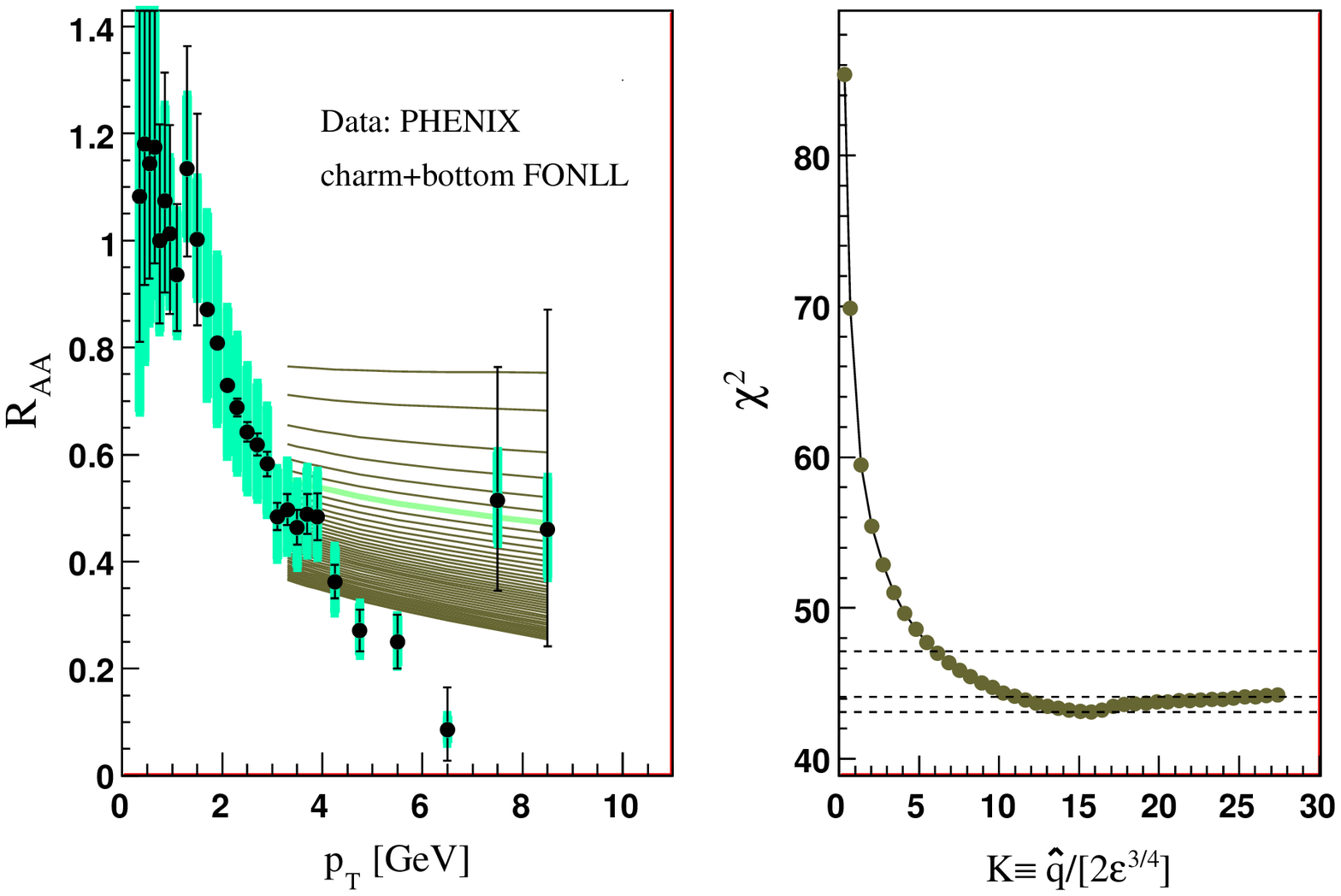}
\includegraphics[width=0.7\textwidth]{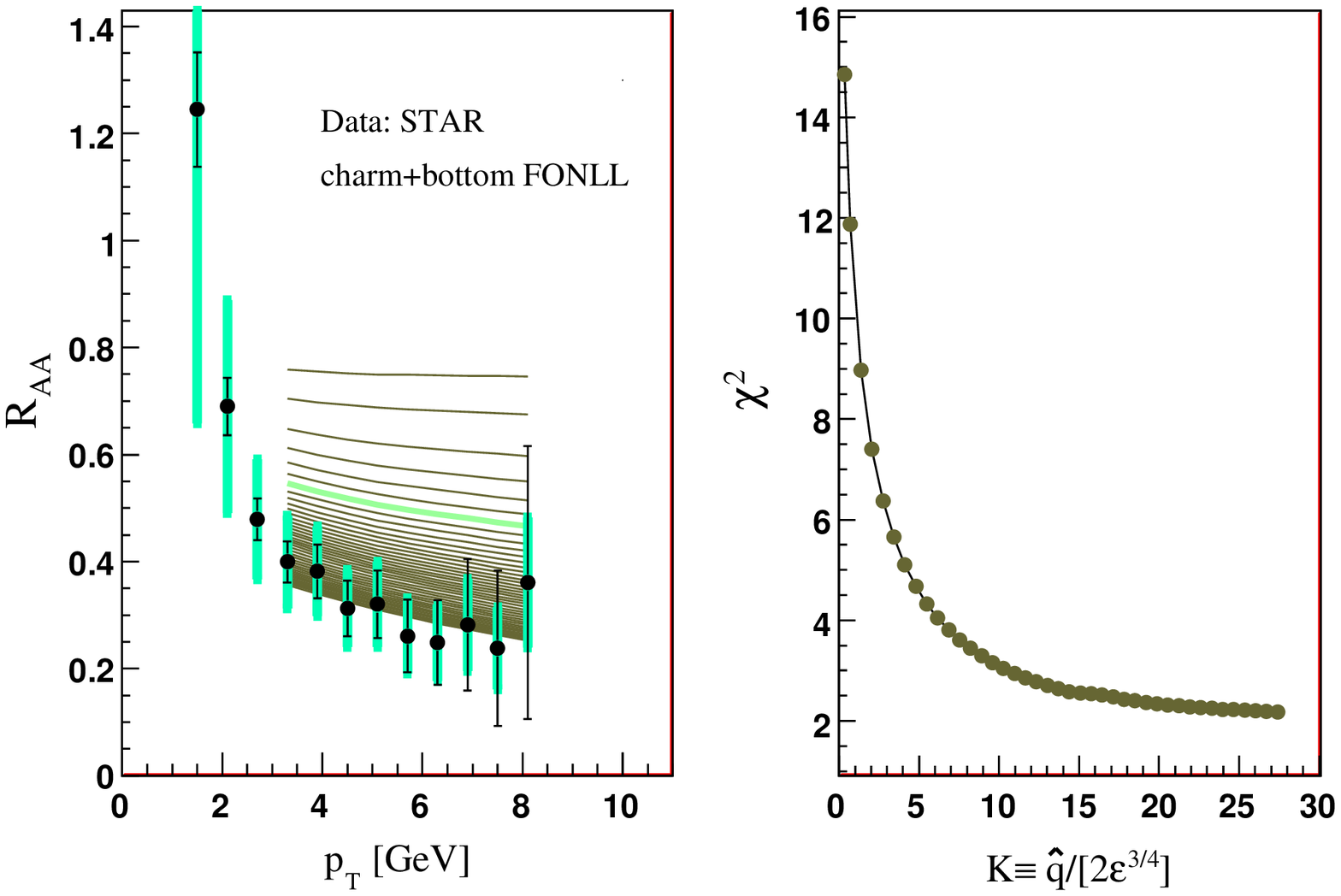}
\end{center}
\caption{Same as Fig. \protect\ref{fig:heavy1} but without nuclear corrections to the PDFs. The horizontal dashed lines in the right panels indicate the minimum of $\tilde\chi^2$ and the corresponding values for one standard deviation, $\Delta\tilde\chi^2=1$, and two standard deviations, $\Delta\tilde\chi^2=4$. Notice that if no local minimum is found in the range of $K$ under study (lower panels corresponding to STAR data) these lines are not shown.}
\label{fig:heavy2}
\end{figure}

\begin{figure}
\begin{center}
\includegraphics[width=0.7\textwidth]{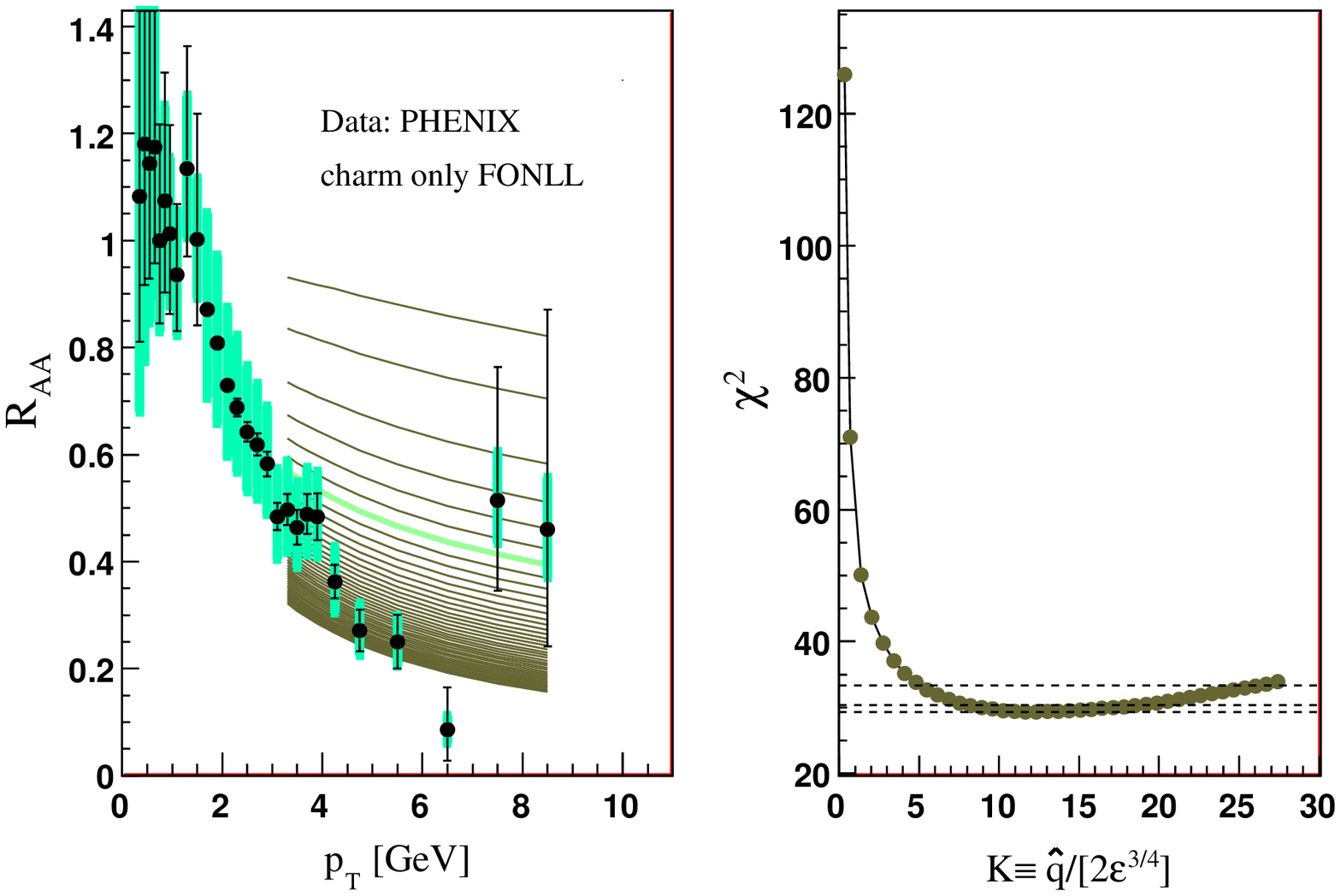}
\includegraphics[width=0.7\textwidth]{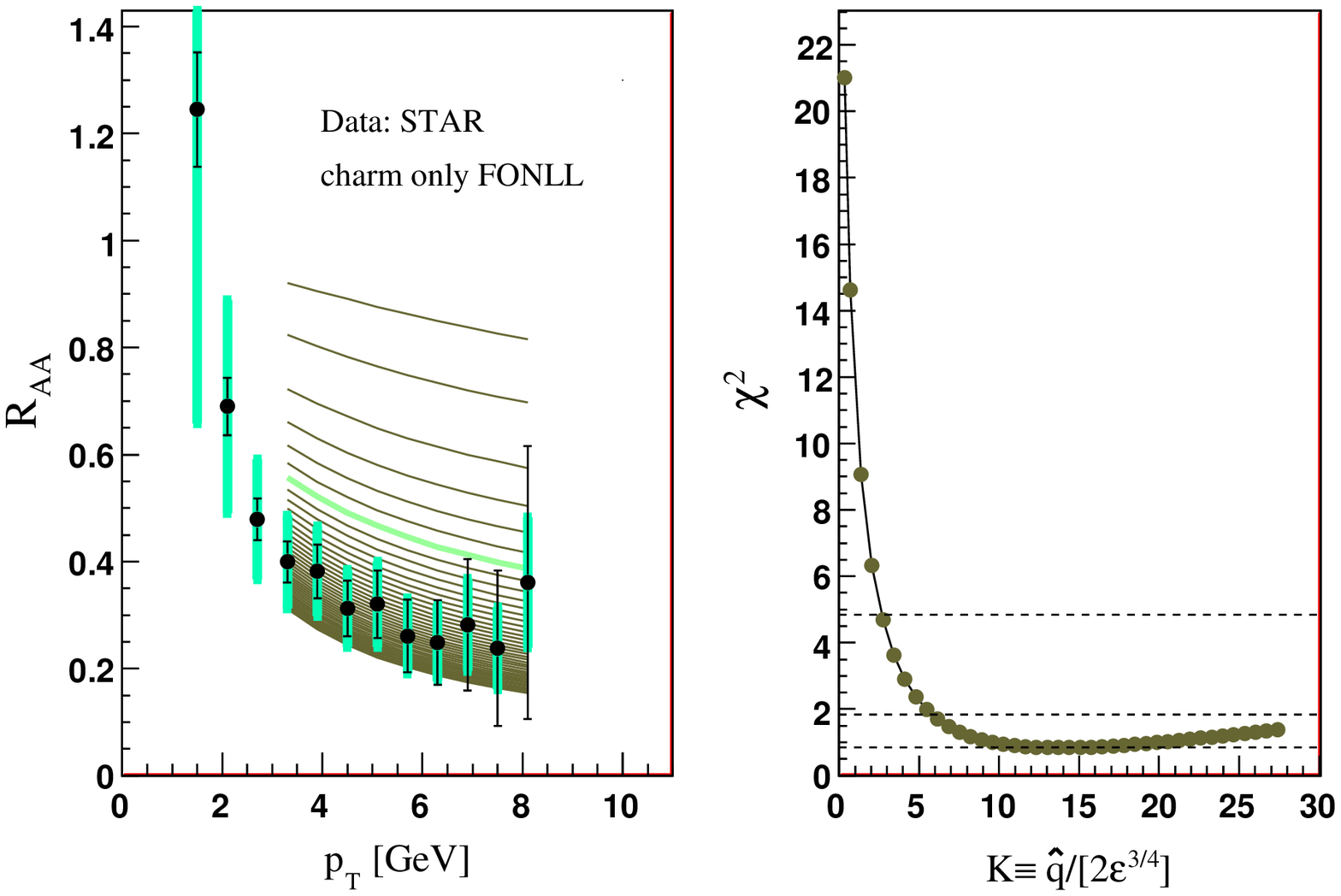}
\end{center}
\caption{Same as Fig. \protect\ref{fig:heavy1} but when only the charm quark is taken into account in the calculation. Horizontal lines correspond to the minimum of $\tilde\chi^2$ and one and two standard deviations as explained in Fig. \protect\ref{fig:heavy2}.}
\label{fig:heavy3}
\end{figure}

In order to further study these data, we make two new computations of the $\tilde \chi^2$ in Fig. \ref{fig:heavy2}
and \ref{fig:heavy3}. 
First we compute the $R_{AA}$ without shadowing corrections. Now the effects are much more visible than for 
light mesons as anti-shadowing is a $20-25\%$ effect here. Second, we compute the suppression when only the 
charm quark contributes to the non-photonic electrons yield, obtaining a much better fit as 
expected --- which would be, in fact, compatible with the determination of $K$ from light mesons. This second observation provides additional support for the need of an experimental identification of the D and B meson suppression.

\section{Summary and conclusions}
\label{conclu}

We have performed a fit to several heavy-ion observables in the large transverse momentum region of the spectrum. 
We combine a realistic hydrodynamical medium profile, obtained from fits to the soft part of the spectrum, with a 
jet quenching calculation within the radiative energy loss formalism in the BDMPS approximation. 
The analysis of the light meson data provides the best fit to $R_{AA}$ and $I_{AA}$ with parameter $K=4.1\pm0.6$, 
for the $K$-factor of the fitted transport coefficient with respect to a perturbative estimate 
from Ref. \cite{Baier:2002tc}. This large value is in qualitative agreement with previous estimates, using 
different medium profiles. This analysis reveals that a common treatment of both one- and two-particle inclusive 
observables leads to statistically compatible results.  An initial investigation of a sub-set of the
theoretical uncertainties reveal the robustness of this fit to reasonable changes of these particular assumptions. 

A similar procedure has been applied to study the case of heavy quarks, as measured via non-photonic electrons.
The baseline spectrum for the heavy quarks has been computed in the FONLL approach. 
Here, the first observation is that the smaller suppression of the formalism, 
when FONLL contributions from charm and bottom are taken into account, leads to a contradiction with the qualitative 
trend of experimental data. The second observation is the large difference of the $\tilde \chi^2$ values of the 
formalism when comparing the theory to the STAR or PHENIX data sets.  Values of the computed 
minimum $\tilde \chi^2$ are very different with the two experimental data sets, 
which make it challenging to make firm statements about 
the statistical agreement between theory and data. 
We have also studied the influence of the shadowing 
corrections in PDFs, and found them to play a more relevant role for heavy quarks compared with the case of 
light mesons. Finally, we have studied the 
case in which all non-photonic electrons come from decays of charmed mesons. In this latter case, the comparison 
gives improved agreement with experimental data due to the smaller mass effect of the charm quark with 
respect to the bottom quark on the radiative energy loss.

A common fit of all these data ('light', 'dihadron', 'heavy') within a single approach and the control 
over the theoretical and experimental uncertainties remains an open issue in jet quenching phenomenology.

\section*{Acknowledgments}
The work of N.A. and C.A.S. has been supported by Ministerio de Ciencia e Innovaci\'on of Spain under projects FPA2005-01963, FPA2008-01177 and contracts Ram\'on y Cajal (NA and CAS), by Xunta de Galicia (Conseller\'{\i}a de Educaci\'on) (NA and CAS) and through grant PGIDIT07PXIB206\-126PR (NA), and by the Spanish Consolider-Ingenio 2010 Programme CPAN (CSD2007-00042) (NA and CAS).
The work of T.H. was partly supported by
Grant-in-Aid for Scientific Research
No.~19740130 and by Sumitomo Foundation
No.~080734. J.L.N. acknowledges funding from the Division of Nuclear Physics of the U.S. Department of Energy under Grant No. DE-FG02-00ER41152.

\end{document}